\documentclass[twoside]{article}
\usepackage{fleqn,espcrc2}

\usepackage{graphicx}

\newcommand{\AmS}{{\protect\the\textfont2
  A\kern-.1667em\lower.5ex\hbox{M}\kern-.125emS}}
\newcommand{\Be}{\begin{equation}}
\newcommand{\Ee}{\end{equation}}
\newcommand{\be}{\begin{displaymath}}
\newcommand{\ee}{\end{displaymath}}
\newcommand{\Ba}{\begin{eqnarray}}
\newcommand{\Ea}{\end{eqnarray}}
\newcommand{\ba}{\begin{eqnarray*}}
\newcommand{\ea}{\end{eqnarray*}}
\newcommand{\Bml}{\begin{mathletters}}
\newcommand{\Eml}{\end{mathletters}}

\newcommand{\om}{\omega}

\newcommand{\LL}{\Lambda_b \rightarrow \Lambda_c}

\begin{document}

\begin{titlepage}

\vspace{-3cm}

\begin{flushright}
MZ-TH/00-43\\
IRB-TH-11/00\\
October, 2000
\end{flushright}

\vspace{0.5cm}
\begin{center}\Large\bf 
Exclusive and Inclusive Semileptonic $\Lambda_b$-Decays
\vspace*{0.3truecm}
\end{center}

\vspace{0.8cm}

\begin{center}
\large J.G.\ K\"{o}rner\\
{\sl Institut f\"{u}r Physik, Johannes Gutenberg-Universit\"{a}t,
     D-55099 Mainz, Germany\\[3pt]} 
\vspace{0.5cm}
and \\
\vspace{0.5cm}
\large Bla\v zenka Meli\'c\\
{\sl Theoretical Physics Division, Rudjer Bo\v skovi\'c Institute,
HR-10002 Zagreb, Croatia\\[3pt]}
\end{center}

\vspace{0.8cm}

\begin{center}
{\bf Abstract}\\[0.3cm]
\end{center}
\parbox{13cm}
\noindent
{In this talk we present theoretical evidence that the exclusive/inclusive ratio
of semileptonic $\Lambda_b$-decays exceeds that of semileptonic
B-decays, where the experimental exclusive/inclusive ratio amounts to about 66\%.
We start from the observation that the spectator quark model provides
a lower bound on the leading order Isgur-Wise function of the $\LL$
transition in terms of the corresponding $ B \rightarrow D,D^* $ mesonic
Isgur-Wise function. Using experimental data for the $ B \rightarrow D,D^* $
mesonic Isgur-Wise functions this bound is established. Applying
a QCD sum rule estimate of
the $\LL$ transition form factor which satisfy the spectator quark model
bound we predict the exclusive/inclusive ratio of semileptonic $\Lambda_b$
decay rates to lie in a range between 0.81 and 0.89.
We also provide an
upper bound on the baryonic Isgur-Wise function which is determined from
the requirement that the exclusive rate should not exceed the inclusive
rate. Our pre-Osaka results are discussed in the light of new recent
preliminary experimental
results on the pertinent mesonic and baryonic form factors presented at the
Osaka ICHEP 2000 Conference.
}

\vspace{2.0cm}

\begin{center}
{\sl Talk given by J.G. K{\"o}rner at the\\
4th International Conference on Hyperons,
Charm and Beauty Hadrons, Valencia, Spain,
27-30 Jun 2000\\
To appear in the Proceeding}
\end{center}

\end{titlepage}
\thispagestyle{empty}
\vbox{}
\newpage

\setcounter{page}{1}



\title{Exclusive and Inclusive Semileptonic $\Lambda_b$-Decays
}

\author{J.G.\ K\"{o}rner
        \thanks{J.G.K. would like to thank the organizers of this conference, 
        Miguel Angel Sanchis Lozano and Jos\'{e} Salt Carols, 
        for providing a most hospitable environment in Valencia.}%
        \address{
        Institut f\"{u}r Physik, Johannes Gutenberg-Universit\"{a}t,
        D-55099 Mainz, Germany}%
        \thanks{e-mail: koerner@thep.physik.uni-mainz.de}
        and
        B. Meli\'{c}\address{
        Theoretical Physics Division, Rudjer Bo\v{s}kovi\'{c} Institute,
        P.O. Box 180, HR-10002 Zagreb, Croatia}%
        \thanks{e-mail: melic@thphys.irb.hr}}
       

\begin{abstract}
In this talk we present theoretical evidence that the exclusive/inclusive ratio
of semileptonic $\Lambda_b$-decays exceeds that of semileptonic
B-decays, where the experimental exclusive/inclusive ratio amounts to about 66\%.
We start from the observation that the spectator quark model provides
a lower bound on the leading order Isgur-Wise function of the $\LL$
transition in terms of the corresponding $ B \rightarrow D,D^* $ mesonic
Isgur-Wise function. Using experimental data for the $ B \rightarrow D,D^* $
mesonic Isgur-Wise functions this bound is established. Applying
a QCD sum rule estimate of
the $\LL$ transition form factor which satisfy the spectator quark model
bound we predict the exclusive/inclusive ratio of semileptonic $\Lambda_b$
decay rates to lie in a range between 0.81 and 0.89.
We also provide an
upper bound on the baryonic Isgur-Wise function which is determined from
the requirement that the exclusive rate should not exceed the inclusive
rate. Our pre-Osaka results are discussed in the light of new recent
preliminary experimental
results on the pertinent mesonic and baryonic form factors presented at the
Osaka ICHEP 2000 Conference.
\vspace{1pc}
\end{abstract}

\maketitle

\section{Introduction}

In mesonic semileptonic $b \rightarrow c$ transitions, the exclusive
transitions to the ground state $S$-wave mesons $B \rightarrow D,D^*$
make up approximately $66 \%$ of the total
semileptonic $B \rightarrow X_c$ rate \cite{PDG}.
In this talk we are concerned about
expectations for the corresponding percentage figure in semileptonic 
bottom baryon decays, i.e. we are 
interested in the ratio of the semileptonic transition rates
$R_E= \Gamma_{\Lambda_b \rightarrow \Lambda_c}/\Gamma_{\Lambda_b
\rightarrow X_c}$. Unfortunately nothing is known experimentally about this
ratio yet. Using some theoretical input and data on bottom meson
decays we predict that the baryonic rate ratio $R_E({\rm baryon})$ lies in
the range
$0.81 \div 0.92$ \cite{KM} and is therefore predicted to be larger
than the corresponding mesonic rate ratio $R_E({\rm meson}) \approx 66 \%$.

In fact this investigation was prompted by two questions on related rate
ratios posted to us by experimentalists. G.~Sciolla asked us about
the semileptonic rate ratio
\Be
R_A=\frac{\Gamma (\LL X l  \nu )}{\Gamma (\Lambda_b \rightarrow X_c l \nu)} ,
\Ee
while P.~Roudeau wanted to know about theoretical expectations for the ratio
\Be
R_B=\frac{\Gamma (\Lambda_b \rightarrow X_c({non}\, \Lambda_c) l \nu )}
{\Gamma (\Lambda_b \rightarrow  X_c l \nu)} .
\Ee
It is very difficult to make reliable theoretical predictions for these
two semi-inclusive and inclusive ratios. However, in as much as one has the
constraint relation
\Be
R_E + R_A + R_B = 1, 
\label{con}
\Ee
and, in as much as all three ratios in (\ref{con}) are positive definite
quantities,
a large number for $R_E$ close to one, as
predicted by us, would limit the ratios $R_A$ and $R_B$ to rather small
values.

The size of the exclusive rate $\Gamma_{\Lambda_b \rightarrow \Lambda_c}$
is tied to the shape of the Isgur-Wise
form factor $F_B(\omega)$ for the $\Lambda_b \rightarrow \Lambda_c$
transition. Expanding $F_B(\omega)$ about the zero recoil point
$\omega=1$, where $F_B(\omega)$ is normalized to one, one writes
\Be
F_B(\om)= F_B(1)[ 1 - \rho_B^2 (\om -1) + c_B\, (\om -1)^2 + ... \,].
\label{exp}
\Ee
The coefficients $\rho_B^2$ and $c_B$ are called the slope parameter and
the convexity parameter, respectively. The slope
parameter $\rho_B^2$ is frequently used to characterize the fall-off
behaviour of the Isgur-Wise function. We have to caution the reader, though,
that it can be quite misleading to use the linear approximation
over the whole range of $\om$ even if the physical range of $(\om -1)$
in this process is
quite small $( 0 \div 0.43)$. For example, if one calculates the rate,
the weight factor which multiplies $F_B(\omega)^2$ in the rate formula is
strongly weighted towards the end of the $\om$-spectrum and one will
thus get quite misleading results if one uses the linear approximation
for the Isgur-Wise function. Besides,
if $\rho_B^2$ exceeds 2.31, $F_B(\omega)$ would become negative
in the physical region which is physically unacceptable.

There is a longstanding controversy about the size of the baryonic
slope parameter
$\rho_B^2$. A first preliminary experimental measurement of
$\rho_B^2$ was presented at the HEP'99 Tampere conference by
the DELPHI Collaboration \cite{DELPHI99}. They obtained the rather large
value of
\Be
\rho_B^2 = 3.4 \pm 1.3 \pm 0.7 .
\label{delphi}
\Ee
Theoretical models offer a wide range of predictions. Taking a
representative set of eight different theoretical models the slope
parameter varies in the range $\rho_B^2= 0.33 \div 2.35$
\cite{EIKL,DHHL,GY,HSM,kroll,KKKK,JMW,SZ}. We emphasize
that this list is not exhaustive. In this talk we present lower and
upper bounds on the slope parameter which read \cite{KM}
\Be
0.36 \leq \rho_B^2 \leq 0.89 \pm 0.19 .
\label{bound}
\Ee
These bounds exclude the model of \cite{EIKL} on the low side and the
models \cite{kroll,KKKK,JMW,SZ} on the high side. Also the preliminary DELPHI
result (\ref{delphi}) can be seen to violate the upper bound.

\section{Origin of bounds}

The upper bound on $\rho_B^2$ has its origin in a spectator quark
model relation which relates
the baryonic form factor to the square of the mesonic form factor. The
relation reads \cite{KKP}
\Be
F_B(\omega) = \frac{\omega + 1}{2} |F_M(\omega)|^2.
\label{spec}
\Ee
The spectator quark model form factor can be seen to provide a lower
bound to the baryonic form factor.
This then leads to an upper bound on the baryon slope parameter given by
\Be
\rho_B^2 \le 2 \rho_M^2 -\frac{1}{2} .
\label{bm}
\Ee
Using an average of the experimental $B \rightarrow D,D^*$ mesonic slope
parameters \cite{Drell}
one then arrives at the upper bound in (\ref{bound}).

The physical picture behind the spectator quark model relation is quite
simple. In the heavy baryon
case there are two light spectator quarks that need to be accelerated
in the current transition compared to the one spectator quark
in the heavy meson transition. Thus the baryonic form factor is determined
in terms of the square
of the mesonic form factor. The factor $(\frac{\omega + 1}{2})$ is a relativistic
factor which insures the correct threshold behaviour of the baryonic
form factor in the crossed
$e^+ e^-$-channel \cite{KKP,IKLR}.

In \cite{IKLR} the relation between heavy meson
and heavy baryon form factors was investigated in the context of
a dynamical Bethe-Salpeter (BS) model. The above spectator quark model
relation (\ref{spec}) in fact emerges when the
interaction between the light quarks in the heavy baryon is switched
off in the BS-interaction kernel. In the more realistic situation when
the light quarks interact with each other,
the heavy baryon form factor becomes flatter, i.e.
the spectator quark model form factor may be used to bound
the heavy baryon form factor from below. This in turn leads to the
upper bound on the slope parameter in (\ref{bm}).

On the other hand, the origin of the lower bound on 
$\rho_B^2$ in (\ref{bound}) derives from
the requirement that the exclusive rate should not exceed the inclusive
rate, i.e.
$\Gamma_{\Lambda_b \rightarrow \Lambda_c} \leq \Gamma_{\Lambda_b
\rightarrow X_c}$ as explained in more detail in the next section.

\section{Numerical values of bounds}

We begin by deriving the lower bound on $\rho_B^2$ in (\ref{bound}). 
As explained before the lower bound is obtained from
the requirement that the exclusive rate
$\Gamma_{\Lambda_b \rightarrow \Lambda_c}$ should not exceed the totally
inclusive rate $\Gamma_{\Lambda_b\rightarrow X_c}$.

The exclusive rate is calculated using the following
input:

\begin{itemize}
\item HQET to $O(1+1/m_Q)$ where the $O(1/m_Q)$ contribution of the
so-called nonlocal form factor $\eta(\omega)$ is dropped. The contribution
of  $\eta(\omega)$ was found to be negligibly small in two model
calculations \cite{DHHL,KKKK}.
\item The $O(1/m_Q^2)$ corrections at zero recoil are fully accounted
for \cite{KP}.
These are extended to the whole $\omega$-range using a technical smoothness
assumption involving the lowest partial wave in the $\LL$-transition. 
\item We use a standard (convex) form of the leading Isgur-Wise function
given by
\Be
F_B(\om) =  \frac{2}{\om +1}\exp \left (-(2 \rho_B^2-1)\frac{\om -1}{\om +1}\right) . 
\label{QCD}
\Ee
which has the correct zero recoil normalization $F_B(1)=1$, a slope
$\rho_B^2$ and a convexity of $(-1+4\rho_B^2+\rho_B^4)/8$. 
\item $O(\alpha_s)$ corrections are included according to the approximate
scheme introduced in \cite{Neubert}.
\end{itemize}

\noindent The inclusive rate $\Gamma_{\Lambda_b\rightarrow X_c}$ is
calculated using the following input:
\begin{itemize}
\item HQET to $O(1+1/m_Q^2)$ thus including the $O(1+1/m_Q^2)$ kinetic
energy correction
\item Full $O(\alpha_s)$ corrections using the results of \cite{Nir}.
\item A pole mass of $m_b=4.8$ GeV from the sum rule calculation of
\cite{PP}
\end{itemize}
Using these ingredients we have obtained
$\Gamma_{\Lambda_b\rightarrow X_c}= 6.50 \cdot 10^{10} s^{-1}$ for the
inclusive rate.

To obtain the numerical value of the lower bound $(\rho_B^2)_{\rm min}=0.36$ we
have adjusted the slope parameter $\rho_B^2$ in the exponential standard form
(\ref{QCD}) such that saturation $R_E=1$ is reached.
As concerns the upper bound we have used the average of the experimental 
values of the mesonic slope parameters in $B \rightarrow D, D^*$
\cite{PDG,Drell} which
we calculate as $\rho_M^2=0.70 \pm 0.10$. This then leads to the upper
bound $\rho_B^2=0.89\pm 0.19$ according to the spectator quark model bound
Eq.(\ref{bm}).

\section{Results on the exclusive/inclusive ratio $R_E$}

We are now in a position to give our results for the exclusive/inclusive
ratio $R_E$. We begin by recording our prediction for the exclusive rate
for which we obtain
\Be
\\ \Gamma_{\Lambda_b \rightarrow \Lambda_c}=5.52 \cdot 10^{10} s^{-1}
\Ee
using $V_{bc}=0.038$. The exclusive rate is calculated using a
slope value of $\rho_B^2=0.75$ which is the average of the two slope
values 0.65 and 0.85 resulting from the QCD sum rule analysis of nondiagonal
and diagonal sum rules, respectively \cite{GY}. This value is identical to
the sum rule result of \cite{DHHL}. We consider the sum rule
calculations to be the most reliable at present. Note that the sum rule
results lie within
the bounds given by Eq.(\ref{bound}).

When calculating
the exclusive/inclusive ratio we allow for a variation of the slope parameter
between these two values of 0.65 and 0.85. Similarly we allow for a
variation of the inclusive rate by using the results of either \cite{PP}
or \cite{HLM}. We thus obtain
\Be
R_E = 0.81 \div 0.92 .
\Ee
Note that the $V_{bc}$-dependence drops out in this ratio. Our conclusion
is that the exclusive/inclusive ratio of semileptonic $\Lambda_b$-decays
is considerably higher than in the corresponding bottom meson case.

\section{Summary}

Let us summarize our findings. Our main predictions are the following:
\begin{itemize}
\item The slope parameter in baryonic $\Lambda_b \rightarrow \Lambda_c$
transitions lies in the range
\Be
0.36 \leq \rho_B^2 \leq 0.89 \pm 0.19 .
\Ee
\item The exclusive rate (using the central value of a QCD sum rule prediction
$\rho_B^2=0.75$ and $V_{bc}=0.038$) is
\Be
\Gamma_{\Lambda_b \rightarrow \Lambda_c}=5.52 \cdot 10^{10} s^{-1}
\Ee
which corresponds to a branching ratio of
$BR(\Lambda_b \rightarrow \Lambda_c l \nu) = 6.8 \%$. Considering the fact that
experimentally one has a semi-exclusive branching ratio of
$BR(\Lambda_b \rightarrow \Lambda_c X l \nu) =
(9.8^{\displaystyle +3.1}_{\displaystyle -3.8})\%$ \cite{PDG} this does not
leave much
room for the inclusive ``X''.
\item The exclusive/inclusive semileptonic rate ratio (using
$\rho_B^2=0.65 \div 0.85$ from QCD sum rules, and \cite{PP} and
\cite{HLM} for $\Gamma_{incl.}$) is predicted to be
\Be
R_E({\rm baryon}) = 0.81 \div 0.92 .
\Ee
$R_E({\rm baryon})$ is thus predicted to be larger than $R_E({\rm meson})
\approx 66\%$.
\end{itemize}

Since the time of this talk two new preliminary experimental results
have appeared that are relevant to the results presented in this
talk. The DELPHI Coll. has come out with new preliminary results on
the slope of the baryonic Isgur-Wise function \cite{DELPHI00}.
They now obtain slope values of
$\rho_B^2 =1.65 \pm 1.3 \pm 0.6$ or, when they include the observed
event rate in the fit, $\rho_B^2 =1.55 \pm 0.60 \pm 0.55$. These new
slope values are considerably smaller than their previous value (\ref{delphi}).
and are now clearly compatible
with the bound (\ref{bound}) even if the central value is still somewhat
high. The results of this new analysis were also presented at this
meeting by T.~Moa \cite{moa}. 

Furthermore, the CLEO Coll. has presented
preliminary
results of a new analysis of the slope parameter in mesonic
$B \rightarrow D^*$-transitions based on a much larger data sample than the one
that was
used in the analysis of \cite{Drell}.
They now obtain $\rho_M^2 = 1.67 \pm 0.11$ \cite{CLEO00}. Using this new
preliminary
value on the mesonic slope parameter would move the upper bound on the
baryonic slope parameter to $(\rho_B^2)_{\rm max} = 2.84 \pm 0.22$.
This new upper bound is much less stringent than the upper bound
$(\rho_B^2)_{\rm max} = 0.89 \pm 0.19$ derived
in \cite{KM} from previous CLEO data. The new upper bound would easily
accommodate all theoretical models mentioned in Sec.I 
(except for \cite{EIKL} which violates
the lower bound) as well as the old
and new measurements of the DELPHI Coll.
\cite{DELPHI99,DELPHI00}. It will be interesting to see whether the
new large mesonic slope value measured by the CLEO Coll. is confirmed by
measurements of BABAR and BELLE which hopefully will become available soon.


\begin{thebibliography}{10}
\bibitem{PDG}
Particle Data Group, C. Caso {\it et al}, Eur. Phys. J. C {\bf 3}, 1 (1998).

\bibitem{KM}
J. G. K{\"o}rner and B. Meli\'{c}, Phys. Rev. D {\bf 62}, 074008 (2000)

\bibitem{DELPHI99}
DELPHI Coll., preliminary paper (DELPHI 99-105 CONF 292) submitted
to the HEP'99 Conference, Tampere, Finland, July 1999.


\bibitem{EIKL}
G.V. Efimov, M.A. Ivanov, N.B. Kulimanova and V.E. Lyubovitskij,
Z. Phys. C {\bf 54}, 349 (1992);
D. Ebert, T. Feldmann, C. Kettner and H. Reinhardt, Z. Phys. C {\bf 71},
329 (1996).

\bibitem{DHHL}
Y.B. Dai, C.S. Huang, M.Q. Huang  and C. Liu, Phys. Lett. B {\bf 387},
379 (1996).

\bibitem{GY}
A.G. Grozin and O.I. Yakovlev, Phys. Lett. B {\bf 285}, 254 (1992).
The numerical evaluation of the sum rules in this published version is
erraneous. The error is corrected in the archive version hep-ph//9908364.
We would like to thank A.G. Grozin for informing us about
the above numerical error.

\bibitem{HSM}
B. Holdom, M. Sutherland and J. Mureika, Phys. Rev. D {\bf 49}, 2359 (1994).

\bibitem{kroll}
M.A. Ivanov, J.G. K{\"o}rner, V.E. Lyubovitskij and P. Kroll,
Phys. Rev. D {\bf 56}, 348 (1997).
\bibitem{KKKK}

B. K{\"o}nig, J.G.  K{\"o}rner, M. Kr{\"a}mer  and P. Kroll, Phys. Rev. D
{\bf 56}, 4282 (1997).

\bibitem{JMW}
E. Jenkins, A. Manohar and M.B. Wise, Nucl. Phys. B {\bf 396}, 38 (1996).

\bibitem{SZ}
M. Sadzikowski and K. Zalewski, Z. Phys. C {\bf 59}, 667 (1993).

\bibitem{KKP}
J. G. K{\"o}rner, M. Kr{\"a}mer and D. Pirjol, Prog. Part. Nucl. Phys.
{\bf 33}, 787 (1994).

\bibitem{Drell}
P.S. Drell, CLNS-97-1521, talk given at 18th International
Symposium on Lepton-Photon Intereactions (LP97),
Hamburg, Germany, 28. Jul - 1. Aug. 1997,
published in "Hamburg 1997 - Lepton-Photon Interactions", p. 347-378.

\bibitem{IKLR}
M.A. Ivanov, J. G. K{\"o}rner, V.E. Lyubovitskij  and A.G. Rusetsky,
Phys. Rev. D {\bf 59}, 074016 (1999).

\bibitem{KP}
J.G. K{\"o}rner and D. Pirjol, Phys. Lett. B {\bf 334}, 399 (1994).

\bibitem{Neubert}
M. Neubert, Nucl. Phys. B {\bf 371}, 149 (1992).

\bibitem{Nir}
Y. Nir, Phys. Lett. B {\bf 221}, 184 (1989).

\bibitem{PP}
A.A. Penin and A.A. Pivovarov, Nucl. Phys. B {\bf 549}, 217 (1999).

\bibitem{HLM}
A.H. Hoang, Z. Ligeti  and A.V. Manohar, Phys. Rev. D {\bf 59}, 074017 (1999).

\bibitem{DELPHI00}
DELPHI Coll., preliminary paper (DELPHI 2000-108 CONF 407)
submitted to the ICHEP00 Conference, July 2000, Osaka, Japan.

\bibitem{moa}
T. Moa, talk given at this conference (to be published in the Proceedings).

\bibitem{CLEO00}
CLEO Coll., preliminary paper (CLEO Conf 00-03; hep-ex/0007052)
submitted to the ICHEP00 Conference, July 2000, Osaka, Japan.

\end{thebibliography}
\end{document}